\documentclass[usenatbib]{mn2e}
\usepackage{psfig}
\title{X-ray bright groups and their galaxies}
\author[S. F. Helsdon \& T. J. Ponman]
        {Stephen F. Helsdon$^{1,2}$\thanks{E-mail: sfh@ociw.edu} and Trevor
         J. Ponman$^{2}$\\
$^1$  The Observatories of the Carnegie Institute of Washington, 813 Santa Barbara Street, Pasadena, CA 91101\\
$^2$  School of Physics and Astronomy, University of
        Birmingham, Edgbaston, Birmingham B15 2TT, UK\\}
 \date{Accepted 2001 ??.
      Received 2001 ??;
      in original form 2001 ??}

\pagerange{\pageref{firstpage}--\pageref{lastpage}}
\pubyear{2001}

\def\Lsol{\hbox{$\thinspace L_{\odot}$}}
\begin{document}

\maketitle

\label{firstpage}

\begin{abstract}
  Combining X-ray data from the ROSAT PSPC and optical data drawn from
  the literature, we examine in detail the relationship between the
  X-ray and optical properties of X-ray bright galaxy groups. We find
  a relationship between optical luminosity and X-ray temperature
  consistent with that expected from self-similar scaling of galaxy
  systems, $L_B \propto T^{1.6\pm0.2}$.  The self-similar form and
  continuity of the $L_B : T$ relation from clusters to groups and the
  limited scatter seen in this relation, implies that the star
  formation efficiency is rather similar in all these systems. We find
  that the bright extended X-ray components associated with many
  central galaxies in groups appear to be more closely related to the
  group than the galaxy itself, and we suggest that these are group
  cooling flows rather than galaxy halos. In addition we find that the
  optical light in these groups appears to be more centrally
  concentrated than the light in clusters.

  We also use the optical and X-ray data to investigate whether early or
  late type galaxies are primarily responsible for preheating in groups.
  Using three different methods, we conclude that spiral galaxies appear to
  play a comparable role to early types in the preheating of the intragroup
  medium. This tends to favour models in which the preheating arises
  primarily from galaxy winds rather than AGN, and implies that spirals
  have played a significant role in the metal enrichment of the intragroup
  medium.

\end{abstract}




\section{Introduction}
\label{sec:intro}

The environment of a galaxy may have a significant effect on the
properties and evolution of the galaxy itself. For example, galaxy
clusters are well known to show a relationship between galaxy
morphology and density (e.g. \citealt{dressler80,dressler97}), which
suggests some sort of environmental influence on the galaxies. By
examining the properties of galaxies as a function of environment it
should be possible to gain important insights into the evolution of
both galaxies and galaxy systems. Particularly interesting is the
group environment, which is typically made up of between three and a
few tens of gravitationally bound galaxies. It is important to
understand the impact of the group environment on galaxies as the
majority of galaxies are found in a group environment \citep{tully87},
and galaxies in clusters and the clusters themselves will have once
been part of groups. In particular, the group environment is likely to
have a significant impact on its galaxies as the density and
velocities of the member galaxies suggest that mergers and
interactions are more common than in clusters or in the field
(e.g. \citealt{mamon92,mamon00}). However, these systems are also
small enough that it is possible for the member galaxies to visibly
affect the properties of the group as a whole.

X-ray studies of groups of galaxies (e.g.
\citealt{helsdon00,mulchaey98b,mahdavi97,ponman96a,mulchaey96,burns96})
provide information about the group environment and the processes
occurring within this environment. Firstly, the presence of diffuse
X-ray emission in a group indicates that the group is likely to be a
real gravitationally bound object, rather than just a chance
superposition of a few galaxies. The X-ray temperature, luminosity and
surface brightness profiles provide information about the depth of the
potential well and the distribution of mass in these systems. In
addition correlations between X-ray parameters can constrain the
effects of other non-gravitational processes in these systems. For
example the relationship between X-ray luminosity and temperature
appears to be steeper in groups than the relation observed in galaxy
clusters \citep{helsdon00,helsdon00b}. This steepening could be the
result of the injection of energy into the intragroup medium by either
starbursts or AGN, and this effect can be reproduced by theoretical
analyses which include the effects of preheating (e.g.
\citealt{balogh99,cavaliere97,cavaliere99}). However, the effects of
cooling may also reproduce this steepening if a large fraction of the
gas cools out within groups
\citep{bryan00,pearce00,voit01,muanwong01}. Preheating of the gas by
the member galaxies is an example of a situation where the galaxies
themselves are affecting the evolution and properties of the group. It
is also likely that the group environment will play a role in shaping
the properties and evolution of the member galaxies.

Little work has been published on the relationship between
X-ray properties of groups and optical properties such as total
optical light and spiral fraction. Previous work (e.g.
\citealt{mahdavi97,mulchaey96}) has mostly been based on samples which are
small or contain optical and X-ray data from a wide variety of sources
which may not be directly comparable with one another. In this and a
companion paper \citep{helsdon02b} we aim to look in some detail at
the relationship between the X-ray and optical properties of X-ray
bright galaxy groups using a consistent approach. In
\citet{helsdon02b} we focus primarily on the morphology-density
relation in X-ray bright groups, whilst here we focus more on the
overall relationship between X-ray and optical group properties. For
the X-ray data we use the 24 galaxy groups analysed in detail by
\citet{helsdon00}. These groups all contain a hot intragroup medium,
and thus represent a sample of bound and collapsed groups. We will
combine this X-ray data with optical data drawn from the literature,
taking care to ensure that the optical data for each group is derived
in a way that enables the groups to be directly compared with one
another. Examining the optical properties for the groups and their
relationship with the X-ray properties will enable us to gain
important insights into the evolution and structure of these systems.

This paper is organised as follows. In \S\ref{sec:sample} we describe
the sample and how we obtain the optical information. We also explain
how we attempt to ensure that the optical data for each group is
derived in a fair and consistent manner. In \S\ref{sec:results} we
present the basic analysis of the data, including relations between
optical properties, such as total optical luminosity, and X-ray
properties such as the group temperature. We also examine the
relationship between the central galaxy and the group, and look at the
morphological makeup of these systems. In \S\ref{sec:dis} we discuss
the implications of our results and look in more detail at some
aspects such as the origin of preheating in these systems. Our
conclusions are summarised in \S\ref{sec:conc}. Throughout this paper
we use $H_0$=50 km s$^{-1}$ Mpc$^{-1}$.


\section{The Sample}
\label{sec:sample}

The X-ray properties of the groups used in this paper are taken from
\citet{helsdon00}, and detailed descriptions of the data reduction and
analysis may be found in that paper. The authors analyse pointed {\it
ROSAT} PSPC observations of 24 X-ray bright groups, and it is these 24
systems which form the sample of groups used in this paper. These
systems were originally identified from three different sources, the
optical group catalogues of \citet{nolthenius93} and \citet{ledlow96}
were examined to identify 15 X-ray bright groups, and then included
were the 9 X-ray bright groups from
\citet{mulchaey98b}. \citet{helsdon00} excluded contaminating sources,
including background point sources and emission from non-central
galaxies, and extracted a count rate and spectrum within a radius
determined for each group by examining a smoothed image and group
profile.  A hot plasma model was then used to obtain a temperature and
derive a bolometric X-ray luminosity after correction for galactic
absorption. This gave 24 groups all with derived luminosities and
temperatures. In addition we carried out 2D fits to the surface
brightness profiles, in order to derive parameters such as
$\beta_{\mathrm{fit}}$, which is effectively a measure of the
steepness of the profile. The $\beta_{\mathrm{fit}}$ values used here
are those from the extended component for the groups with 2 component
models in \citet{helsdon00}, and that of a single component for the
remainder.

When deriving the optical properties of each group (total $L_B$,
spiral fraction, etc) it is not satisfactory to simply use the galaxy
memberships as given in the original group catalogue, since the
galaxies were originally selected from different sources which in
general have different selection criteria. Ideally all galaxies in
each group down to the same absolute magnitude would be included. Thus
we initially define two different magnitude cuts, these are the
magnitudes at which 50\% and 90\% of the total light in an assumed
luminosity function would be included.  The luminosity function used
in this case is that derived from deep imaging and spectroscopy by
\citet{zabludoff00} for X-ray bright groups, which is a Schechter
function with $\alpha = -1.3$ and $M_* = -23.1$. This luminosity
function applies in the $R$-band, whereas throughout this paper $B$-band
magnitudes are used. To convert to $B$-band magnitudes, colours given in
\citet{fukugita95} are used, giving $B-R=1.57$ for early-type
galaxies. We use the $B-R$ for early types, as the majority of the
galaxies in these X-ray bright groups are early-types.  This procedure
gives an absolute magnitude cut in the $B$-band of $M_B=-20.55$ ($\sim
0.4 L_*$) for the 50\% cut, and $M_B=-16.32$ ($\sim 0.008 L_*$) for
the 90\% cut. For the `typical' group in our sample
(v$\approx$5000km/s) this corresponds to an apparent magnitude of
$\sim$14.4 for the 50\% cut and $\sim$17.2 for the 90\% cut.

For each group we searched the NASA/IPAC Extragalactic Database (NED)
for galaxies lying within the group virial radius in projection on the
sky, and having recession velocities within three times the group
velocity dispersion (3$\sigma_{\rm g}$) from the catalogued group
mean. The virial radius of each group (typically $\sim$ 1.1 Mpc) is
calculated using $R_V~=~1.14\,(T/1\,\rm keV)^{\frac{1}{2}}$ Mpc which
can be derived from relations obtained in simulations by
\citet{navarro95a}. Whilst it is known that the velocity dispersion of
a group may be significantly underestimated when based on only a few
members (e.g. \citealt{zabludoff98}) this should not be a big problem
for this sample. Over half the sample have velocity dispersions
calculated from 15 members or more, and all but 5 have velocity
dispersions calculated from at least 7 members. The centre position
used for each group is the centre of the X-ray emission. In most cases
this position is very close to the position of the central galaxy,
with the three exceptions being bimodal systems in which the X-ray
centre falls roughly between the two main galaxies.

\begin{table*}
\begin{center}
\small
\begin{tabular}{@{\extracolsep{-0.2cm}}lccccccccccc}
Group        & $L_X$            & $T$             & $\beta_{\textrm{fit}}$   & total $L_{\textrm{opt}}$     & early $L_{\textrm{opt}}$     & late $L_{\textrm{opt}}$       & $f_{\textrm{sp}}$ & $f_{\textrm{sp}}$ & $\Delta m_{12}$ & $N_{gal}$ & $\sigma_v$ \\
name         & log(erg s$^{-1}$)& (keV)           &                 & (\Lsol)             &(\Lsol)              & (\Lsol)              & num.     & light    & (mag.)&    & (km s$^{-1}$)     \\
\hline
NGC 315      & 42.15 $\pm$ 0.15 & 0.85 $\pm$ 0.07 & 1.37$\pm$0.36   & 2.94$\times10^{11}$ & 2.27$\times10^{11}$ & 6.69$\times10^{10}$  & 0.60     & 0.23     & 1.47  &  5 & 122 \\ 
NGC 383      & 43.31 $\pm$ 0.02 & 1.53 $\pm$ 0.07 & 0.48$\pm$0.02   & 8.02$\times10^{11}$ & 5.93$\times10^{11}$ & 1.46$\times10^{11}$  & 0.21     & 0.18     & 0.00  & 35 & 466 \\  
NGC 524      & 41.37 $\pm$ 0.11 & 0.56 $\pm$ 0.08 & 0.45$\pm$0.01   & 1.99$\times10^{11}$ & 1.65$\times10^{11}$ & 3.37$\times10^{10}$  & 0.36     & 0.17     & 2.25  & 11 & 205 \\ 
NGC 533      & 42.95 $\pm$ 0.02 & 1.06 $\pm$ 0.04 & 0.43$\pm$0.03   & 6.03$\times10^{11}$ & 2.45$\times10^{11}$ & 3.59$\times10^{11}$  & 0.50     & 0.59     & 0.14  & 16 & 464 \\ 
NGC 741      & 42.66 $\pm$ 0.03 & 1.08 $\pm$ 0.06 & 0.391$\pm$0.009 & 4.49$\times10^{11}$ & 3.50$\times10^{11}$ & 9.96$\times10^{10}$  & 0.53     & 0.22     & 2.27  & 19 & 432 \\ 
NGC 1587     & 41.50 $\pm$ 0.18 & 0.92 $\pm$ 0.15 & 0.47$\pm$0.06   & 2.27$\times10^{11}$ & 1.18$\times10^{11}$ & 9.96$\times10^{10}$  & 0.50     & 0.44     & 0.10  & 7  & 106 \\ 
NGC 2563     & 42.79 $\pm$ 0.02 & 1.06 $\pm$ 0.04 & 0.400$\pm$0.004 & 5.31$\times10^{11}$ & 2.84$\times10^{11}$ & 2.46$\times10^{11}$  & 0.50     & 0.46     & 0.57  & 20 & 336 \\ 
NGC 3091$^1$ & 42.20 $\pm$ 0.03 & 0.71 $\pm$ 0.03 & 0.41$\pm$0.02   & 3.63$\times10^{11}$ & 3.25$\times10^{11}$ & 3.79$\times10^{10}$  & 0.25     & 0.10     & 1.58  & 12 & 211 \\ 
NGC 3607     & 41.59 $\pm$ 0.03 & 0.41 $\pm$ 0.04 & 0.45$\pm$0.04   & 1.97$\times10^{11}$ & 1.74$\times10^{11}$ & 2.33$\times10^{10}$  & 0.40     & 0.12     & 0.88  & 10 & 421 \\ 
NGC 3665     & 41.36 $\pm$ 0.10 & 0.45 $\pm$ 0.11 & 0.49$\pm$0.03   & 1.15$\times10^{11}$ & 1.09$\times10^{11}$ & 6.38$\times10^{9}$   & 0.33     & 0.06     & 1.33  & 3  & 29 \\ 
NGC 4065     & 42.99 $\pm$ 0.04 & 1.22 $\pm$ 0.08 & 0.41$\pm$0.01   & 9.39$\times10^{11}$ & 6.77$\times10^{11}$ & 2.62$\times10^{11}$  & 0.33     & 0.28     & 0.37  & 18 & 495 \\ 
NGC 4073     & 43.70 $\pm$ 0.01 & 1.59 $\pm$ 0.06 & 0.46$\pm$0.01   & 8.69$\times10^{11}$ & 7.03$\times10^{11}$ & 1.13$\times10^{11}$  & 0.20     & 0.13     & 1.65  & 23 & 607 \\ 
NGC 4261     & 42.32 $\pm$ 0.02 & 0.94 $\pm$ 0.03 & 0.35$\pm$0.03   & 1.18$\times10^{12}$ & 5.39$\times10^{11}$ & 6.42$\times10^{11}$  & 0.49     & 0.54     & 1.23  & 57 & 465 \\ 
NGC 4325     & 43.35 $\pm$ 0.03 & 0.86 $\pm$ 0.03 & 0.60$\pm$0.01   & 2.13$\times10^{11}$ & 1.48$\times10^{11}$ & 5.13$\times10^{10}$  & 0.14     & 0.24     & 0.46  & 11 & 256 \\ 
NGC 4636     & 42.48 $\pm$ 0.01 & 0.72 $\pm$ 0.01 & 0.373$\pm$0.008 & 4.00$\times10^{11}$ & 2.81$\times10^{11}$ & 1.18$\times10^{11}$  & 0.47     & 0.30     & 0.07  & 17 & 463 \\ 
NGC 4761$^2$ & 43.16 $\pm$ 0.01 & 1.04 $\pm$ 0.02 & 0.364$\pm$0.006 & 6.38$\times10^{11}$ & 4.04$\times10^{11}$ & 2.20$\times10^{11}$  & 0.45     & 0.35     & 0.05  & 26 & 376 \\ 
NGC 5129     & 42.78 $\pm$ 0.04 & 0.81 $\pm$ 0.06 & 0.44$\pm$0.01   & 5.60$\times10^{11}$ & 3.57$\times10^{11}$ & 1.96$\times10^{11}$  & 0.70     & 0.35     & 0.91  & 12 & 294 \\ 
NGC 5171     & 42.92 $\pm$ 0.05 & 1.05 $\pm$ 0.11 & 0.34$\pm$0.03   & 4.32$\times10^{11}$ & 1.83$\times10^{11}$ & 6.27$\times10^{10}$  & 0.40     & 0.15     & 0.89  & 16 & 424 \\ 
NGC 5353$^3$ & 41.76 $\pm$ 0.03 & 0.68 $\pm$ 0.05 & 0.44$\pm$0.02   & 5.72$\times10^{11}$ & 1.94$\times10^{11}$ & 3.76$\times10^{11}$  & 0.67     & 0.66     & 0.64  & 16 & 174 \\ 
NGC 5846     & 42.36 $\pm$ 0.02 & 0.70 $\pm$ 0.02 & 0.58$\pm$0.01   & 4.07$\times10^{11}$ & 3.28$\times10^{11}$ & 7.85$\times10^{10}$  & 0.29     & 0.19     & 0.40  & 14 & 368 \\ 
NGC 6338     & 43.93 $\pm$ 0.01 & 1.69 $\pm$ 0.16 & 0.52$\pm$0.04   & 5.97$\times10^{11}$ & 5.35$\times10^{11}$ & 4.40$\times10^{10}$  & 0.29     & 0.07     & 0.34  & 8  & 587 \\ 
NGC 7176$^4$ & 41.47 $\pm$ 0.11 & 0.53 $\pm$ 0.11 & 1.07$\pm$0.29   & 1.82$\times10^{11}$ & 9.81$\times10^{10}$ & 8.37$\times10^{10}$  & 0.59     & 0.46     & 0.51  & 17 & 193 \\ 
NGC 7619     & 42.62 $\pm$ 0.02 & 1.00 $\pm$ 0.03 & 0.78$\pm$0.08   & 3.61$\times10^{11}$ & 2.91$\times10^{11}$ & 6.27$\times10^{10}$  & 0.50     & 0.17     & 0.06  & 20 & 253 \\ 
NGC 7777     & 41.75 $\pm$ 0.20 & 0.62 $\pm$ 0.15 & 0.35$\pm$0.02   & 1.45$\times10^{11}$ & 1.44$\times10^{11}$ & 0                    & 0.00     & 0.00     & 0.05  & 3  & 116 \\ 
\hline
\end{tabular}
\normalsize
\end{center} 
\caption{\label{tab:data}The X-ray and optical data. $L_X$, $T$ and
  $\beta_{\textrm{fit}}$ are the X-ray luminosity (bolometric and
  absorption corrected), temperature and extended component surface
  brightness index from \protect\citet{helsdon00}; total
  $L_{\textrm{opt}}$, early $L_{\textrm{opt}}$ and late
  $L_{\textrm{opt}}$ are the total optical luminosity, early-type
  luminosity and late-type luminosity all in the $B$-band. The columns
  titled $f_{\textrm{sp}}$ are the spiral number fraction and spiral
  light fraction respectively. $\Delta m_{12}$ is the difference in
  magnitude between the first and second ranked galaxy. $N_{gal}$ is
  the number of galaxies to the 90\% luminosity cut within the virial
  radius and $\sigma_v$ is the group velocity dispersion as given in
  \protect\citet{helsdon00}. The flagged group names have the
  following commonly used alternative names --- $^1$--HCG 42,
  $^2$--HCG 62, $^3$--HCG 68, $^4$--HCG 90. }
\end{table*}

We also only selected galaxies brighter than the groups 90\% apparent
magnitude cut.  Unfortunately this selection alone gives no indication
of how complete the sample of galaxies are in each group. To try to
better determine this, the online Lyon-Meudon Extragalactic Database
(LEDA) catalogue was searched over the same position and velocity
range. This search found no extra galaxies to each group's 50\%
magnitude cut, and only 7 out of the 24 groups had galaxies added to
the 90\% cut (typically only increasing the total optical luminosity
by $\sim$ 4\%). For galaxies with a known redshift the LEDA catalogue
is $\approx$ 90\% complete to a $B$-band magnitude of 14.5
\citep{amendola97,paturel97}.  The typical group in this sample has the
50\% magnitude cut at $\sim$14.4 so most groups should be at least
90\% complete to their 50\% cut, just from the LEDA samples. Given
this, and the fact that NED almost always lists more galaxies than
LEDA (many of these systems have been the subject of specific
membership studies which have been added to NED) we believe that our
derived optical properties (to our 90\% cut) such as the total $B$-band
luminosity, while an underestimate, will be at the worst a factor of
two too low, with most cases better than this. An estimate of the
typical incompleteness can be made by comparing over all the groups
the total optical luminosity within the 90\% cut with 1.8 (=0.9/0.5)
times the total luminosity within the 50\% cut. This comparison
suggests that typically we are missing ~35\% of the total group light
at the 90\% cut. However, given that there could be some trends in
incompleteness in the optical data, we will consider below the
possible effects of incompleteness on our results.

Morphological types are taken firstly from NED if available, and then
LEDA.  The galaxies are split into early (elliptical and S0) and late
(spiral and irregular) type samples. Galaxies on the borders of the
morphological classes are put into the earlier type --- e.g. an E-S0
will be classed as elliptical and an S0a will be classed as an S0. For
each group we calculate the total light, light in early types, light
in late types, fraction of light in late types and the difference in
magnitude between the first and second ranked galaxies. The galaxy
magnitudes used from NED are those listed on the initial search
page. This means that the magnitudes of the individual galaxies may
not all be exactly comparable to one another, but the effect of this
on the derived total group optical luminosities should be small
compared to the effects of incompleteness.

\begin{figure*}
\hspace{0cm}
\psfig{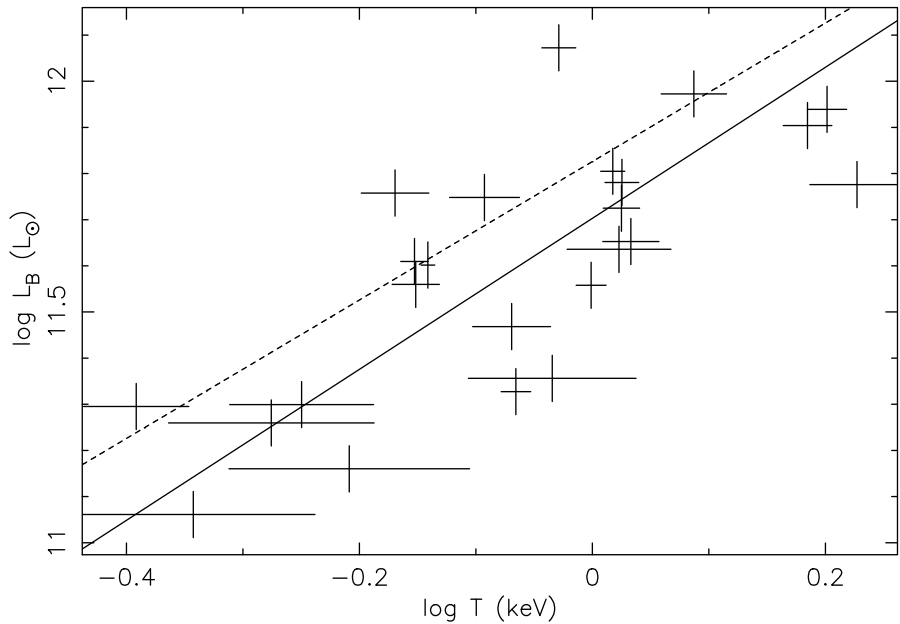}
\caption{\label{fig:temp_allopt}Group optical luminosity versus X-ray
  temperature. Solid line is best fit to data, and dashed line is the
  joint fit to groups and clusters by \protect\citet{lloyddavies00b}.}
\end{figure*}

The derived optical properties for each group are listed in
table~\ref{tab:data} along with the appropriate X-ray derived
parameters. All optical parameters are derived for the virial radius,
3$\sigma_g$ sample based on the 90\% magnitude cut, using the
distances given for the groups in \citet{helsdon00}. It can be seen
that in a few cases the early type light and late type light does not
add up to the total light. This is because these groups have some
galaxies which did not have types listed in either NED or LEDA,
although the contribution to the total light from these untyped
galaxies is almost always much less (all $<$10\% apart from NGC 5171)
than the contribution from galaxies with known morphological type.

It should be noted that the group sample used here should not be
regarded as being statistically complete in any way. However, we do
not believe that this will introduce any particular bias, other than
the fact that since we only use groups with detected diffuse X-ray
emission, we do not include systems with undetectably faint
intergalactic gas. The group sample should rather be regarded as a
reasonably representative sample of X-ray bright groups.


\section{Results}
\label{sec:results}

\subsection{Optical light in groups}
\label{sec:opt_light_results}

The X-ray temperature of a galaxy group is a measure of the depth of
the gravitational potential well, and hence a measure of the system
size and mass. It is clearly of interest then, to examine how the
temperature relates to the total optical light in a group. Thus we
plot in Figure~\ref{fig:temp_allopt} the total $B$-band light for each
of the groups as a function of X-ray temperature. It is clear that
there is a significant correlation between these two parameters, and
the significance of Kendall's rank correlation coefficient (a
distribution-free test for correlation) is $K=3.77$ ($P=0.00017$ of
chance occurrence). Any trend in optical incompleteness, if present,
would have the effect of producing a correlation in the opposite
direction to that observed here (the hotter systems tend to be further
away). As far as the authors are aware, this is the first time that a
correlation between these two parameters has been reported for galaxy
groups. Given the origin of the galaxy magnitudes it is difficult to
calculate an exact error on the integrated $B$-band luminosity for
every group and errors on the log $L_B$ values are all taken to be
$\pm$ 0.05. Comparison of this value with typical errors of galaxy
magnitudes in NED suggest that this is a reasonable estimate of the
error. The distribution of the points (with known temperature errors)
suggest that there is more scatter than would be expected due to
statistical errors alone. As a result of this, we do not weight by
statistical errors when fitting regression lines to the data.  All
lines are derived using the program SLOPES \citep{feigelson92} which
derives six different linear regressions on the data, along with error
analysis. All fits given here use the bisector of the ordinary least
squares regression of $y$ on $x$ and $x$ on $y$, as this method
performs best for a symmetrical treatment of the variables
\citep{isobe90}.  Regression uncertainties are derived using jackknife
resampling as the standard formulae for calculating uncertainties
underestimate the errors for $N \leq$ 50 \citep{feigelson92}.

The best fit line to the data presented in Figure~\ref{fig:temp_allopt} is\\

\noindent
\begin{math}
\log L_B = (11.70 \pm 0.04) + (1.64 \pm 0.23) \log T
\end{math}
.
\\

Also plotted for comparison is the line derived by
\citet{lloyddavies00b} for a sample of systems ranging from groups to
clusters. These authors attempted to correct for the effect of any
incompleteness in their optical sample, and derived a line with a
slope of 1.5 $\pm$ 0.2. As can be seen, the slopes of the two
relations are consistent with one another, and the lower normalization
of the present sample is consistent with the Lloyd-Davies \& Ponman
line if 25\% to 35\% of the optical light is missed in the low
luminosity end of the luminosity function for our groups. This is
consistent with what we would expect given our sample selection as
described in
\S\ref{sec:sample}.

\begin{figure*}
\hspace{0cm}
\psfig{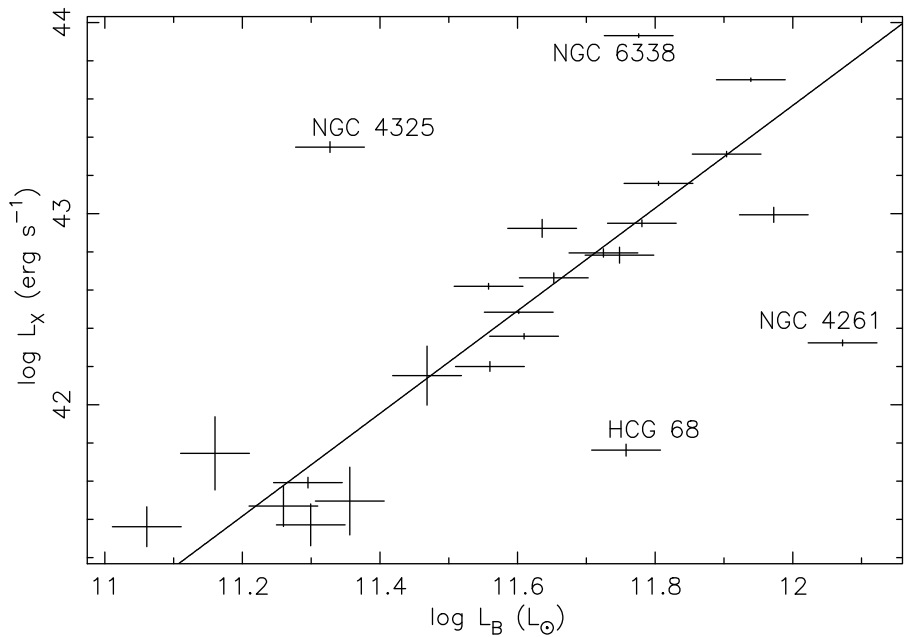}
\caption{\label{fig:lx_lb}Group X-ray luminosity versus group optical 
  luminosity. Solid line is best fit to data.}
\end{figure*}

\begin{figure*}
\hspace{0cm}
\psfig{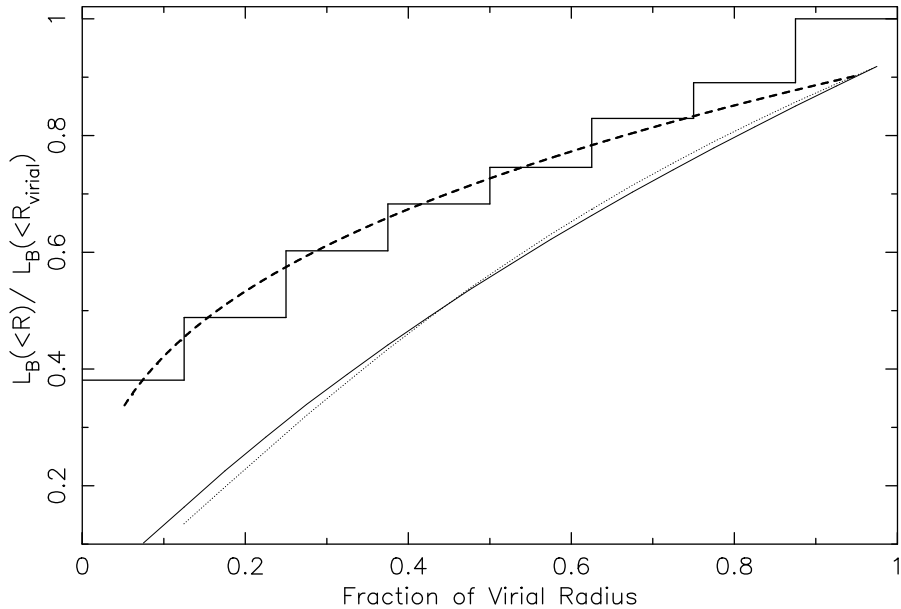}
\caption{\label{fig:lumrat}The fraction of the total cumulative
light versus projected radius (in units of the virial radius), for all
groups in this sample. The bold dashed line is the best fit
powerlaw. The dotted line is the equivalent line for clusters inferred
from the \protect\citet{carlberg97} Hernquist model fit to the surface
number density profile derived from their sample of 14 clusters, and
the faint solid line is the profile inferred from the
\protect\citet{carlberg01} CNOC2 group surface number density profile.}
\end{figure*}

\begin{figure}
\hspace{0cm}
\psfig{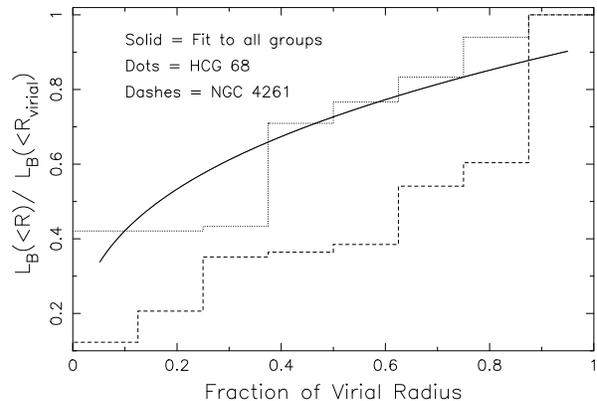}
\caption{\label{fig:twogrp}The fraction of the total cumulative
light as a function of projected radius (in units of the virial
radius), for HCG~68 (dotted line) and NGC~4261 (dashed line). For
comparison the best fit powerlaw to all groups is also shown.}
\end{figure}

Given this correlation between $L_B$ and temperature, and the known
correlation between X-ray luminosity and temperature
\citep{helsdon00,helsdon00b}, it is expected that there would also be a
correlation between $L_X$ and $L_B$. In Figure~\ref{fig:lx_lb} we plot
$L_X$ versus $L_B$ for the 24 groups in this sample. As can be seen
there is a significant correlation ($K=3.82$, $P=0.00014$). Once
again, any systematic trend in incompleteness would act to produce a
correlation in the opposite sense to that observed - the more X-ray
luminous groups tend to be further away and are thus more likely to be
incomplete. We fit a line using the method outlined above (ignoring
the four marked outliers which are discussed below) to obtain,\\

\noindent
\begin{math}
\log L_X = ( 40.88 \pm 0.18 ) + (2.69 \pm 0.29) \log \left ({\textstyle L_B\over \textstyle 10^{11}\,L_\odot }\right)
\end{math}
,
\\

\noindent which is also plotted in Figure~\ref{fig:lx_lb}.
Alternatively, we could have plotted $L_X/L_B$ vs $L_B$ which gives a
best fit line of $\log L_X/L_B = (29.83\pm0.18)+(1.77\pm0.28)\log
(L_B/10^{11}\,L_\odot)$. Note that we express these two fits in x-axis units
of $L_B/10^{11}\,L_\odot$ to avoid a large correlation between the errors in
slope and intercept. While the line appears to generally describe the
trend well, there are four points which deviate significantly (note
that the inclusion of these four points does not significantly alter
the best fit --- including these points gives an intercept and
gradient of $40.91\pm0.21$ and $2.63\pm0.36$ respectively). The first
of these outliers, NGC~4325, lies well above the trend, and is
probably unusually X-ray luminous as it also lies well above the X-ray
$L_X : T$ relation. The origin of this high X-ray luminosity is
unclear -- as far as we are aware, no indications have been reported
of AGN activity, and the galaxy is undetected in the FIRST
\citep{white97first} and NVSS radio surveys
\citep{condon98}. The other group lying above the trend, NGC~6338 (the 
least severe of these outliers), may have some contamination by an AGN
\citep{hwang99}, although it is also the most distant system in this 
sample, and thus its optical luminosity may be underestimated more
than the other groups. The remaining two groups, HCG~68 and NGC~4261
fall well below the trend, and although they both do fall just below
the group $L_X : T$ relation, they do not have a particularly low
$L_X$ for their temperature. The alternative is that these two systems
may have an excess of optical light, possibly due to luminous galaxies
at the borders of the systems. To investigate this we plot in
Figure~\ref{fig:lumrat} the fraction of the total cumulative light as
a function of group virial radius, for all groups in this sample. In
Figure~\ref{fig:twogrp} we show the corresponding plot for HCG~68 and
NGC~4261 separately. 

As can be seen in Figure~\ref{fig:lumrat} these groups typically
contain $\sim60\%$ of their total light within a projected distance of
a third of their virial radius. We have fit a powerlaw to these data
and find a best fit of $L_B(<r) \propto r^{0.34\pm0.06}$ (last data
point was ignored as a few groups showed `kicks' in their profiles in
the final bin --- see below). For comparison, galaxy clusters tend to
have projected galaxy number densities $\propto r^{\alpha}$ with
$-1.4\leq{\alpha}\leq-1.0$ (e.g.
\citealt{beers86,oegerle87,squires96,carlberg97,trevese00}). This 
implies projected cumulative light profiles in clusters of between
$L_B(<r) \propto r^{0.6}$ and $L_B(<r) \propto r^{1.0}$, consistent
with that actually observed in some clusters
(e.g. \citealt{squires96}). To demonstrate this more clearly, we show
the cluster profile inferred from the \citet{carlberg97} Hernquist
model fit to the surface density profile derived from their sample of
14 clusters (the normalisation is adjusted to match our fit). This
means that clusters appear to have much steeper cumulative light
profiles, suggesting that the optical light (and galaxies --- the
implied group 3D galaxy density profile is $\propto r^{-2.66}$) in
X-ray bright groups is more centrally concentrated than in typical
clusters.

This distribution of optical light is more concentrated than would be
inferred from previous measures of the galaxy distribution in loose
groups \citep[][also displayed in
Figure~\ref{fig:lumrat}]{hickson88b,walke89,carlberg01}, and less
concentrated than similar estimates for Hickson compact groups
\citep{mendes94,montoya96}. However these differences are not
particularly surprising. Previous studies of loose groups are based on
optically selected samples, and it is possible that they included some
spurious groups or systems at an early stage of virialisation (only a
fraction are likely to be X-ray bright), both of which could easily
act to produce a less concentrated profile. Additionally, in almost
all systems in this X-ray bright group sample, a bright early-type
galaxy is located at the centre, whereas the centres of optically
defined groups may not be located on any galaxy, which again would act
to produce a less concentrated profile. In comparison, the compact
groups, while likely real systems, have generally only been studied to
a small fraction of the radii used here, and large extrapolations are
needed, on the profiles which are not strongly constrained.

As for offsets from the group $L_X:L_B$ relation,
Figure~\ref{fig:twogrp} clearly shows that NGC~4261 has an unusual
light profile. In fact, almost half the group optical light is found
beyond a radius of 0.75~R$_{\rm{V}}$. Clearly, for this group its high
optical luminosity is caused by galaxies at large radius which might
only now be joining the group, or might even be nearby foreground or
background objects --- this group is near Virgo. Whilst a few other
groups show similar `kicks' at large radius, NGC~4261 is the most
extreme example by far. In contrast, the remaining $L_X:L_B$ outlier,
HCG~68, does not show much evidence of contamination at large radius
and just appears to have a very high optical luminosity relative to
its X-ray luminosity. It should be noted that although these plots
show evidence of a small amount of contamination at large radii in
these groups, this contamination only represents a small fraction of
the total light (5-10\%), an even smaller fraction of the total number
of galaxies and as such we do not expect it to significantly affect
our main results.

\begin{table*}
\begin{center}
\begin{tabular}{lllcc}
Relation                               & correlation & P       & intercept      & slope \\
\hline
Total optical light versus temperature & 3.77        & 0.00017 & 11.70 $\pm$ 0.04 & 1.64 $\pm$ 0.23 \\
Early type light versus temperature    & 3.52        & 0.00043 & 11.52 $\pm$ 0.04 & 1.53 $\pm$ 0.19 \\
Late type light versus temperature     & 1.82        & 0.069 & 11.11 $\pm$ 0.11 & 2.3 $\pm$ 0.7 \\
 & & & & \\
(Total optical light/$10^{11}$) versus $L_X$       & 3.82        & 0.00014 & 40.88 $\pm$ 0.18 & 2.69 $\pm$ 0.29 \\
(Early type light/$10^{11}$) versus $L_X$          & 3.77        & 0.00017 & 41.26 $\pm$ 0.15 & 2.87 $\pm$ 0.31 \\
(Late type light/$10^{11}$) versus $L_X$           & 1.29        & 0.2     & 42.54 $\pm$ 0.17 & 1.36 $\pm$ 0.24 \\
\hline
\end{tabular}
\end{center}  
\caption{\label{tab:morph1}Correlations and fits to full and morphological
  subsets of data versus X-ray temperature and luminosity. All fits are in
  log space.}
\end{table*}

\begin{figure}
\hspace{0cm}
\psfig{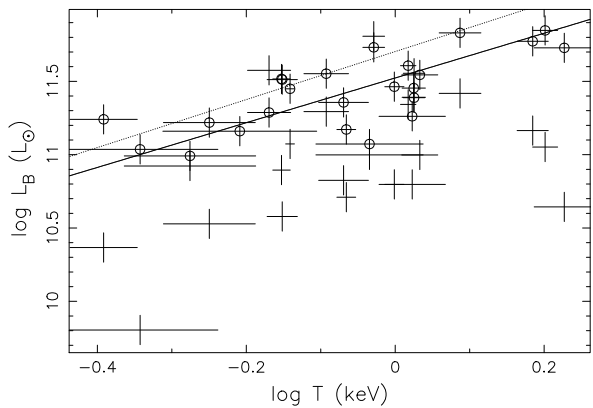}
\caption{\label{fig:morph2}Group optical luminosity in early type 
(circled crosses) and late type (plain crosses) galaxies versus group
gas temperature. Dotted line is the best fit line from the total
optical light data. Solid line is best fit to early-type points.}
\end{figure}

\begin{figure}
\hspace{0cm}
\psfig{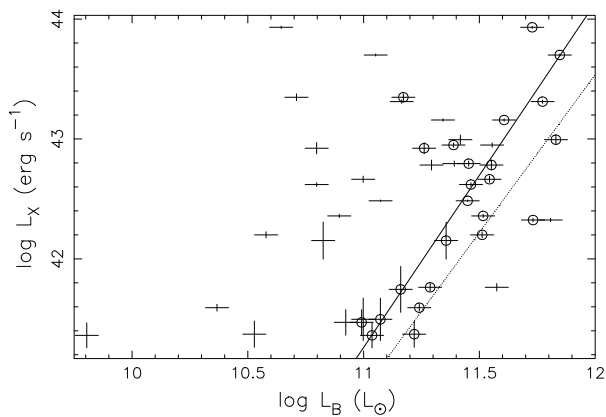}
\caption{\label{fig:morph3}Group X-ray luminosity versus group 
 optical luminosity in early type (circled crosses) and late type
 (plain crosses) galaxies respectively. Dotted line is the best fit
 line from the total optical light data. Solid line is best fit to
 early-type points.}
\end{figure}

\begin{figure*}
\hspace{0cm}
\psfig{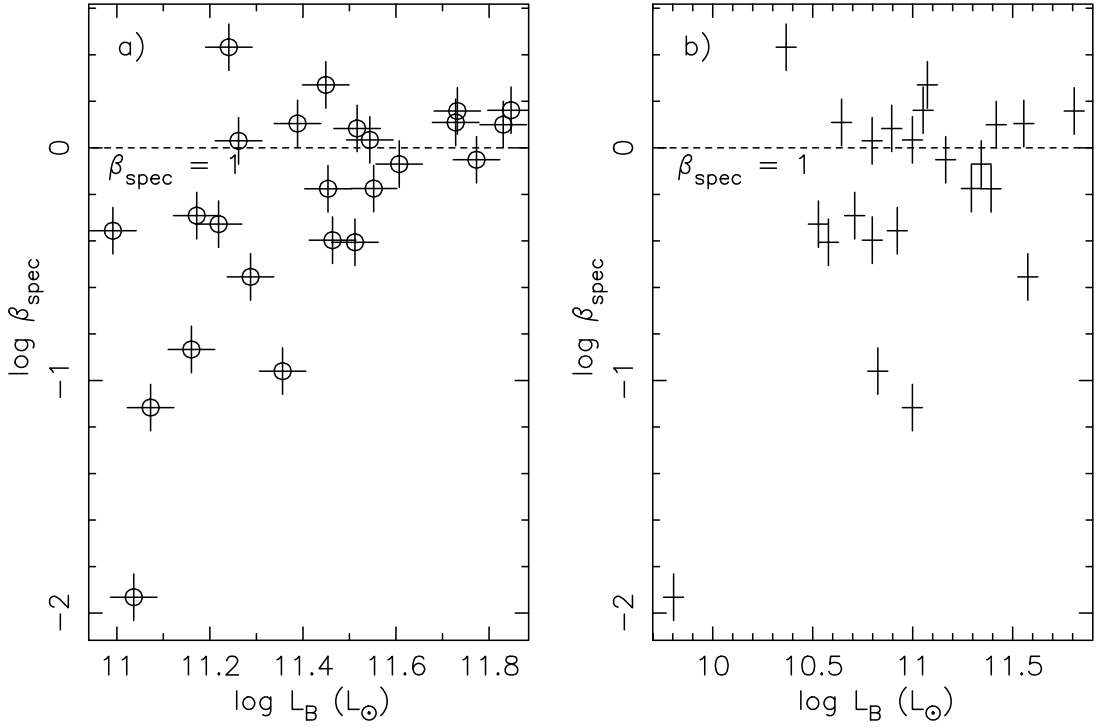}
\caption{\label{fig:betas_both}Group $\beta_{\textrm{spec}}$ versus total 
  early type light (a) and total late type light (b). The dashed line
  corresponds to $\beta_{\textrm{spec}}=1$.}
\end{figure*}

We have also examined the $L_B : T$ and $L_X : L_B$ relations as a function 
of morphological type. The data for these are plotted in
Figures~\ref{fig:morph2} and \ref{fig:morph3} and the correlation
strengths and fit results are summarised in table~\ref{tab:morph1}. As
can be seen the total light in early type galaxies correlates well
with both X-ray temperature and luminosity whilst the same relations
are only weakly correlated for the late types. Another interesting
parameter which shows a difference when plotted between the total
early and late type light is $\beta_{\textrm{spec}}$, which is derived from the
group velocity dispersion and temperature ($\beta_{\textrm{spec}} = \mu
m_p\sigma_{v}^2/kT$), and is a measure of the ratio of the specific
energy in the galaxies to the specific energy in the gas. This is
shown in Figure~\ref{fig:betas_both}.

As can be seen in Figure~\ref{fig:betas_both} there is a significant
correlation between $\beta_{\textrm{spec}}$ and the light in
early-types ($K=2.98$, $P=0.003$), but not with the late-type light
($K=1.08$, $P=0.28$). There is unlikely to be any trend due to
incompleteness, as $\beta_{\textrm{spec}}$ does not correlate with
distance. Low values of $\beta_{\textrm{spec}}$ would tend to suggest
that there is more energy in the gas than in the galaxies and that
some sort of energy has been added to the gas. A naive interpretation
of this result would be that late type galaxies do not contribute in a
systematic way to the injection of energy, and that systems with few
early types have had the most energy injected. However there are a
number of complicating factors. Smaller systems will show the effects
of any energy injection more readily than larger systems, and will
also tend to contain a lower fraction of early-types (e.g. see
Figure~\ref{fig:spfr_vs_temp}). Most of the light in these systems is
contained in early-type galaxies, so there is likely to be more
scatter in relations involving late-type light. Finally
$\beta_{\textrm{spec}}$ is calculated from the group velocity
dispersion which may be quite poorly determined for some of the poorer
systems in this sample. We will return to the issue of the origin of
injection of energy into the intragroup medium in
\S\ref{sec:dis:pre}.

\subsection{Central galaxies in Groups}

Most X-ray bright groups contain a bright early-type galaxy located at
the centroid of the group X-ray emission. These galaxies are also
centrally located according to both the projected and velocity
distributions of the member galaxies \citep{zabludoff98}, and may be
formed by galaxy merging which occurs soon after the initial collapse
of the group \citep{governato96}. Given their special position within
the group these galaxies are likely to have properties closely related
to the group. For example, the position angle of the optical light of
the central galaxy tends to align with the position angle of the group
X-ray emission \citep{mulchaey98b}. In a previous paper
\citep{helsdon01} we derived X-ray luminosities for 11 central galaxies in
groups for which we had two-component models. These two-component
models allowed the separation of a component coincident with the
central galaxy and an extended component associated with the bulk of
the group. Here we examine the relationship between the X-ray and
optical properties of these central galaxies and those of the group as
a whole.

\begin{figure*}
\hspace{0cm}
\psfig{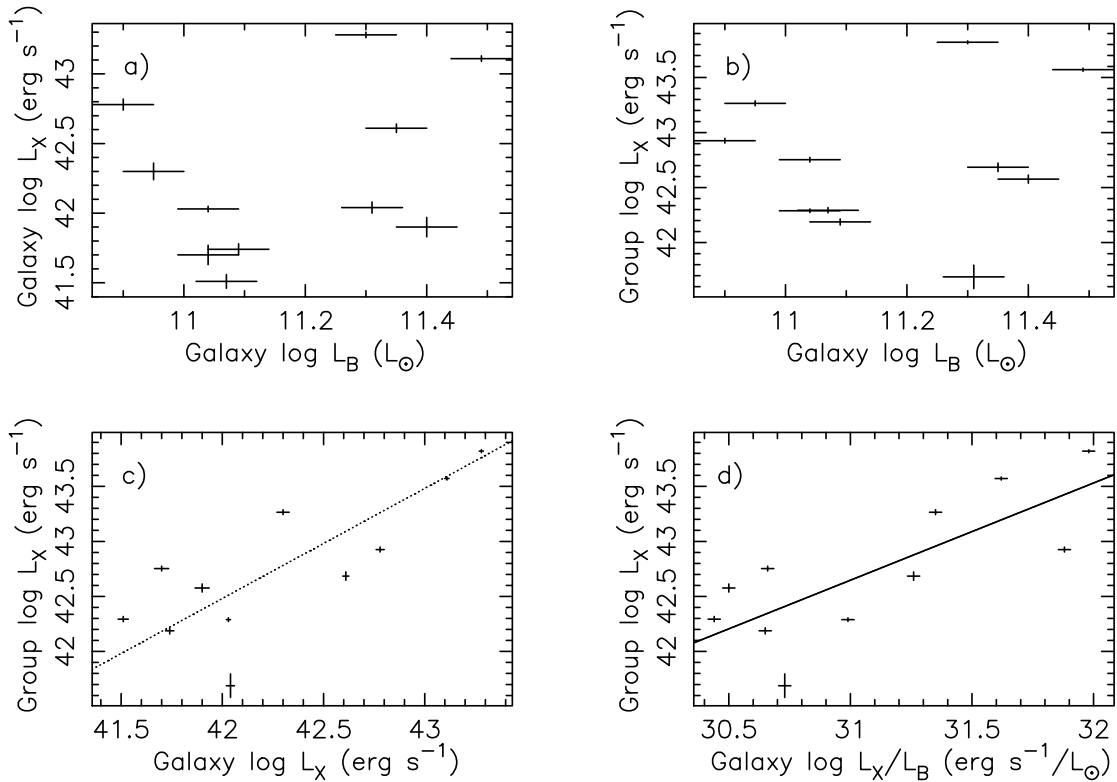}
\caption{\label{fig:cen}a) Central galaxy $L_X : L_B$
  relation. The dashed line is from \protect\citet{beuing99}. b) Group
  $L_X$ versus galaxy $L_B$. c) Galaxy $L_X$ versus group
  $L_X$. Dotted line marks galaxy $L_X$ = 33\% of group $L_X$. d)
  Group $L_X$ versus galaxy $L_X/L_B$. Solid line is best fit to
  data. The group luminosities do not include the central galaxies.}
\end{figure*}

In Figure~\ref{fig:cen}, we plot 4 graphs showing the relationship
between central galaxy X-ray luminosity (luminosity of central X-ray
component), central galaxy optical luminosity, and group X-ray
luminosity (for the 11 groups with two component models to the X-ray
emission). We use a slightly different group X-ray luminosity in
Figure~\ref{fig:cen} as the luminosities quoted in \citet{helsdon00}
(and used throughout the rest of this paper) include the contribution
assigned to the central galaxy. We subtract this central component
from the group luminosity to avoid any bias when comparing the group
luminosity to the galaxy X-ray luminosity (which is calculated from
this central component). Firstly we show the galaxy $L_X : L_B$ for
these galaxies, and plot for comparison the best fit line obtained by
\citet{beuing99} for a large sample of early-types (in a variety of
different environments). It can be seen that the points are not
strongly correlated ($K=0.47$) and that they scatter about the
\citet{beuing99} line. This is not particularly surprising as the
range in $L_B$ is small and the early-type $L_X : L_B$ relation is
known to have a scatter of between one and two orders of magnitude (
e.g. \citealt{canizares87,eskridge95} ). We also show the relationship
between central galaxy optical luminosity and group X-ray luminosity,
which shows no correlation ($K=-0.31$). However there is a pronounced
correlation between the central galaxy X-ray luminosity and the group
X-ray luminosity ($K=2.10$, $P=0.036$). This correlation is
approximately linear with the X-ray luminosity of the central galaxy
being about 33\% that of the group (shown by the dotted line ---
equivalent to 25\% of the total group luminosity including this
component). The correlation is strongest in the final plot which shows
the group $L_X$ against the galaxy $L_X/L_B$ ($K=2.41$, $P=0.016$),
along with the best fit line (slope=0.9 $\pm$ 0.2).

The much stronger correlation between the X-ray properties of the
central galaxies with the group, rather than the optical light of the
galaxy suggests that the majority of the X-ray emission of these
central X-ray components is associated with the group, rather than the
galaxy itself. Given that almost all of the groups shown in
Figure~\ref{fig:cen} show evidence of a temperature drop in their
central regions \citep{helsdon00}, it seems likely that the central
components seen in these X-ray bright groups is actually due to a
group cooling flow, rather than any property of the galaxy itself.

\begin{figure}
\hspace{0cm}
\psfig{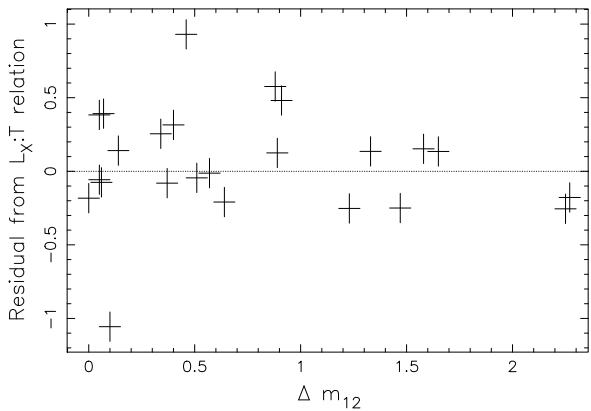}
\caption{\label{fig:ltres_d12}Residual from group $L_X : T$ relation as a 
  function of $\Delta m_{12}$. Dotted line marks zero residual from
  relation.}
\end{figure}

Another interesting issue to examine is whether systems which have a
very dominant central galaxy (i.e. much brighter than any other group
members) show any difference from systems with several galaxies of
similar optical luminosity. The dominance of the central galaxy is
estimated by obtaining the difference in blue magnitude between the
brightest and second brightest galaxies in the group ($\Delta
m_{12}$). Strictly speaking, $\Delta m_{12}$ only measures the
dominance of the central galaxy if the central galaxy is indeed the
brightest galaxy. For two of the groups (NGC 5353 and NGC 4261) the
central galaxy is not the brightest galaxy and both have a more
luminous spiral in the outer regions. For consistency we simply use
the definition of $\Delta m_{12}$ above but note that excluding the
bright spiral in these groups does not significantly change our
results.  The brightest galaxy in all other groups is a centrally
located early-type.  It is of particular interest to compare $\Delta
m_{12}$ with offset from the mean group $L_X : T$ relation. It might be
expected that systems with high values of $\Delta m_{12}$ will lie
above the relation, since there is some evidence that fossil groups
\citep{ponman94,mulchaey99a,vikhlinin99,jones00}, which generally have
$\Delta m_{12} \geq 2.5$ lie above the $L_X : T$ relation \citep{jones00}.

In Figure~\ref{fig:ltres_d12} we plot $\Delta m_{12}$ versus residual
from the $L_X : T$ relation given by \citet{helsdon00b}. As can be seen,
there is no trend in the data, and no suggestion that systems with
high values of $\Delta m_{12}$ lie above the $L_X : T$ relation. We also
find no correlation of $\Delta m_{12}$ with any other group property.

\subsection{Spiral fractions}

X-ray bright groups tend to have low spiral fractions \citep{mulchaey96}
and generally contain bright early-type galaxies at their centres
\citep{zabludoff98}. Given that there is some sort of connection between
the morphological makeup of a group and its X-ray properties, it is
natural to look for trends in X-ray properties as a function of spiral
fraction.

\begin{figure*}
\hspace{0cm}
\psfig{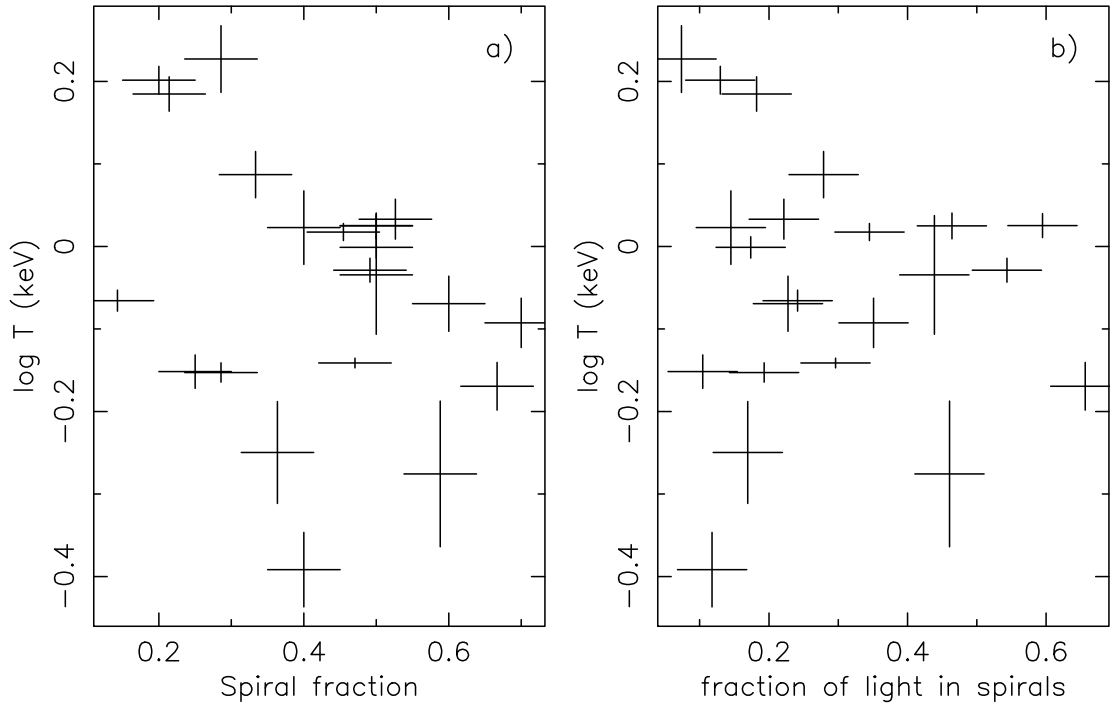}
\caption{\label{fig:spfr_vs_temp}X-ray temperature of galaxy groups as a
  function of both a) spiral number fraction and b) the fraction of
  light in spirals. Note that the two groups with less than four typed
  members have been excluded.}
\end{figure*}

We initially examine the relation between spiral fraction and group
temperature. In Figure~\ref{fig:spfr_vs_temp} we plot the group
temperature as a function of spiral fraction and fraction of light in
spirals. Note that only 22 systems are plotted, as we have excluded 2
systems with fewer than 4 galaxies of known morphological type. There
is a weak trend ($K=-1.3$) for cooler systems to contain more spirals,
although the fraction of light contained in spirals shows a lot more
scatter ($K=-0.54$). These trends are consistent with those seen in
the Hickson compact groups \citep{ponman96a}, but offset relative to
those seen in clusters \citep{edge91b}. This offset is in the sense
that for a particular spiral fraction the groups have a much lower
temperature relative to the cluster trend. This is consistent with a
scenario in which the morphological transformation of spirals to
ellipticals is more efficient in the group environment as is suggested
from the morphology-density relation of these systems
\citep{helsdon02b}. We have also examined residuals from the group
$L_X : T$ relation to check if any of the scatter in this relation was
related to the morphology of the galaxies present. However we found no
trend with spiral fraction (e.g. see Figure~\ref{fig:ltfr} later) or
with the fraction of total group light in spirals.

\begin{figure*}
\hspace{0cm}
\psfig{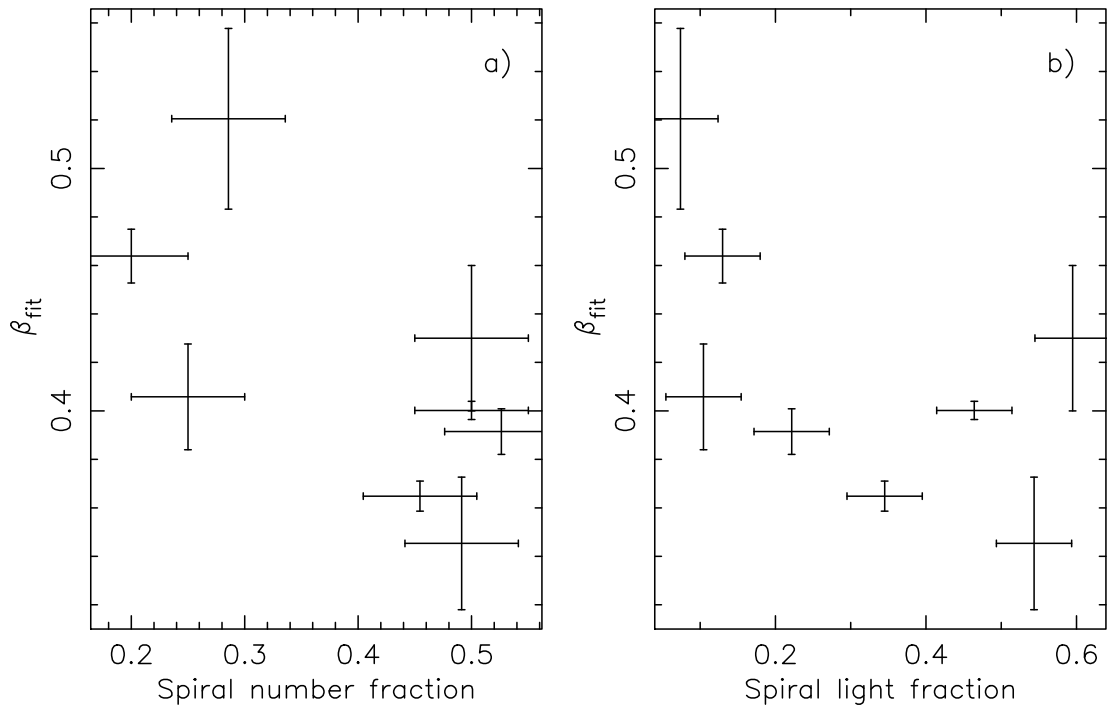}
\caption{\label{fig:beta_spiral}$\beta_{\mathrm{fit}}$ as a function of a) spiral
  number fraction and b) spiral light fraction, for the eight groups
  with the most reliable values of $\beta_{\mathrm{fit}}$ identified
  in \protect\citet{helsdon00}.}
\end{figure*}

Finally we have examined trends in $\beta_{\mathrm{fit}}$ as a
function of both spiral number fraction and spiral light
fraction. $\beta_{\mathrm{fit}}$ is derived by modelling the surface
brightness distribution of the systems with a modified King profile
($\beta$-model) of the form
$S(r)=S_0(1+(r/a)^2)^{-3\beta_{\mathrm{fit}}+0.5}$, where the surface
brightness as a function of radius, $S(r)$, is determined by the
central surface brightness, $S_0$, the core radius, $a$ and the index,
$\beta_{\mathrm{fit}}$.  Thus, $\beta_{\mathrm{fit}}$ is effectively a
measure of the slope of the gas profile in these groups, with low
values indicating a flatter profile. Although the
$\beta_{\mathrm{fit}}$ values for these systems are derived from
systems with emission observed to different fractions for the group
virial radius, this is unlikely to introduce significant biases in the
$\beta_{\mathrm{fit}}$ values. This is because the core radii of these
systems are generally much smaller than the maximum extent of the
observed emission \citep{helsdon00}. Furthermore, \citet{sanderson02}
show that, at least for clusters, truncating the profiles does not
introduce any significant bias in the derived values of
$\beta_{\mathrm{fit}}$.

Unfortunately, $\beta_{\mathrm{fit}}$ is still a difficult parameter
to measure, especially in data with poor statistics. In addition it is
likely that most groups' surface brightness profiles are described by
a two-component model \citep{helsdon00}, whereas the data are often
only of sufficient quality to fit a single component model.  Given
this, it is no surprise that the plots of $\beta_{\mathrm{fit}}$
versus spiral fraction for the full sample show a lot of scatter and
no significant correlation. However \citet{helsdon00} identified a
subset of eight groups which have good quality data, and which are
well fit by two-component models.  $\beta_{\mathrm{fit}}$ for the
extended component in each of these eight systems is plotted as a
function of each of the spiral fractions in
Figure~\ref{fig:beta_spiral} (We use the alternative value of
$\beta_{\mathrm{fit}}$=0.43 for NGC~533 for the reasons discussed in
\citealt{helsdon00}). While it is clear that there is still real
scatter in these plots, both the spiral number fraction and the spiral
light fraction show a weak negative trend ($K=-1.13$ and $K=-1.48$
respectively, probability of chance occurrence, $P=0.258$ and
$P=0.139$). This trend is in the sense that groups with more spirals
have flatter profiles. It is possible that these trends in
$\beta_{\mathrm{fit}}$ as a function of spiral fraction could just be
the result of trends in $\beta_{\mathrm{fit}}$ and spiral fraction as
a function of mass (more massive systems tend to have higher values of
$\beta_{\mathrm{fit}}$ and lower spiral fractions). This will be
discussed in more detail in
\S\ref{sec:dis:pre}.


\section{Discussion}
\label{sec:dis}

\subsection{Optical light in groups and its relationship to the X-ray
  properties} 

It has been shown in the previous section that the optical properties
of X-ray bright galaxy groups are quite strongly related to their
X-ray properties (Figure~\ref{fig:temp_allopt}). Firstly we have seen
that the total optical light in these systems appears to correlate
well with both X-ray temperature and luminosity. The slope of the
relation between optical luminosity and temperature, $L_B \propto
T^{1.6 \pm 0.2}$, agrees well with the relation predicted from
self-similar scaling of galaxy systems, $L_B
\propto T^{1.5}$ (assuming $L_B$ traces mass), and with that derived
from galaxy clusters ($L_B \propto T^{1.5 \pm 0.2}$,
\citealt{lloyddavies00b}).

The fact that both groups and clusters appear to lie on the same
self-similar $L_B : T$ relation is interesting as it has important
implications for the star formation efficiency (i.e. the fraction of
the baryonic mass found in stars) in these systems.  Some previous
work has suggested that there is a trend of increasing star formation
efficiency from clusters to groups
\citep{david90,arnaud92,pildis95,bryan00}. However if this were the
case then one would also expect the $L_B : T$ relation for groups to
be flatter and/or have a higher normalisation than the self-similar
cluster relation. Note also, that while using $L_B$ as a trace of
system mass could introduce a small bias (lower temperature systems
tend to have a higher fraction of spirals which could give an
enhancement in $L_B$ relative to the true stellar mass), removing this
bias would tend to steepen the observed relation in the direction of
lower star formation efficiency in groups.  Therefore, the apparent
accordance of the $L_B : T$ relation with the self-similar trend seen
in clusters suggests that the star formation efficiency does not
change significantly between X-ray bright groups and clusters,
although it could still change if other parameters conspired to hide
this effect. For example, the star formation efficiency would be best
quantified from the available mass in gas and the unique $L_B : T$
relation could hide a gas mass fraction that increases with
temperature and an optical luminosity to gas mass ratio drops with
increasing temperature.

An apparently constant star formation efficiency agrees with recent work by
\citet{lloyddavies00b} who calculate the star formation efficiency for a
number of groups and clusters and see no apparent trend, and with
theoretical work by \citet{baugh99} who predict approximately the same
fraction of cold baryons in group and cluster sized halos. The
contrary results from earlier observational work most likely arises
because values were derived at a small fraction of the virial radius
and extrapolated using a $\beta$-model with a value of
$\beta_{\mathrm{fit}}$=0.67.  Galaxy groups tend to have flatter X-ray
profiles than this \citep{helsdon00}, and will therefore contain a
large fraction of their gas mass at large radii.  Using a value of
$\beta_{\mathrm{fit}}$=0.67 would tend to underestimate the gas mass
in groups which, in turn, would lead to an overestimate of the star
formation efficiency.

A constant star formation efficiency in groups and clusters has
implications for models which invoke cooling as a possible cause of
entropy injection \citep{pearce00,voit01,muanwong01}. In these models
radiative cooling results in the removal of low entropy gas, which in
turn leads to an increase in temperature and reduction in luminosity,
which is most noticeable in groups. However, if the star formation
efficiency is indeed the same in clusters and groups then any cooling
gas could not form large quantities of stellar material as this would
act to increase the star formation efficiency. Instead the material
would have to cool to some dark form.

We have also examined the scatter about the $L_B : T$ relation, as there
is clearly more scatter than the statistical errors alone would
suggest. A Monte Carlo approach was used to quantify the amount of
scatter observed beyond that expected from the statistical errors
alone. Initially the orthogonal deviation of the points about the best
fit line was calculated.  This gave a measure of the total scatter,
given that scatter in both the $x$ and $y$ directions may be
important. To calculate the scatter expected from the statistical
errors we took a dataset in which all the points lay on the best fit
line and we then scattered the points in the $x$ and $y$ direction
according to the statistical errors. This was repeated 1000 times to
obtain an average orthogonal deviation expected from the statistical
errors. By subtracting this expected statistical scatter from the
total scatter, a measure of the intrinsic real orthogonal scatter is
obtained. By assuming that all of this orthogonal scatter has its
origin in either the $x$ or $y$ direction we can estimate the maximum
possible scatter in each direction.

For the $L_B : T$ relation we find that there is about twice as much
scatter about the best fit line than the statistical errors would
suggest.  This corresponds to a scatter of 45\% in $L_B$ for accurate
$T$ or a scatter of 26\% in $T$ for accurate $L_B$. The 45\% scatter
allowable in $L_B$ is actually fairly small, given that some of this
scatter is likely to be in the temperature, and that there may be
varying levels of optical completeness amongst the groups which would
act to increase the scatter.  For comparison, the non-statistical
scatter in $L_X$ for a fixed $T$ is more than 130\%.  This small
scatter in $L_B$ suggests that variations in the star formation
efficiency of X-ray bright galaxy groups must also be small
($<$45\%). Thus overall it would appear that all X-ray bright groups
have a fairly similar star formation efficiency, which is similar to
that for galaxy clusters.

The relation obtained between X-ray and total optical luminosity
(figure~\ref{fig:lx_lb}), $L_X \propto L_B^{2.6 \pm 0.4}$, is
consistent with the relation expected given the previously derived
$L_B : T$ relation and the known $L_X : T$ relation
\citep{helsdon00b}. This relation has been examined previously by
\citet{mahdavi97} for groups in the CfA redshift survey, who derive a much
flatter relation of $L_X \propto L_B^{1.06 \pm 0.11}$. However
\citet{mahdavi97} only have a total of nine data points after removing
upper limits, and their X-ray luminosities are based on RASS data, rather
than the better quality pointed data used here. Furthermore
\citet{mahdavi97} extract their X-ray data from a region defined by the
optical positions of the galaxies, thus they also include any emission from
the galaxy members themselves. This means that their flatter relation may
be caused by the galaxy contribution (not necessarily related to the group)
becoming significant at lower luminosities.

We have also examined the scatter about the $L_X : L_B$ relation, as for the
$L_B : T$ relation, (after removing the 3 discrepant points -- see
\S~\ref{sec:opt_light_results}) and find that there is $\sim$2 times as
much scatter as the statistical errors would suggest. Taking the
scatter in each of the two directions separately we find that the
scatter in $L_B$ is comparable to that derived for the $L_B : T$
relation, whilst the scatter in $L_X$ is $\sim$100\%, which is
comparable to the scatter seen in $L_X$ about the $L_X : T$
relation. Thus, as a function of temperature the scatter in $L_X$ is
much greater (3 to 4 times larger) than the scatter in $L_B$.  This
suggests that physical processes which vary from group to group have a
more marked effect on the gas properties than on the galaxies. For
example, processes which have moved gas around will have a significant
effect on $L_X$, as $L_X$ is dependent on the square of the gas
density (e.g. the groups may have experienced variable amounts of
preheating). It is also worth noting that the self-similar slope of
the $L_B : T$ relation suggests that T has been little affected by any
preheating (since any boost in T for low temperature groups would be
expected to steepen the $L_B : T$ relation), whilst in contrast, both
the relations involving $L_X$ ($L_X : T$ and $L_X : L_B$) show a
significant steepening, indicating that $L_X$ has been reduced in cool
systems.

Differences in the $L_B : T$ and $L_X : L_B$ relations as a function
of morphological type are interesting but care is needed in their
interpretation. At a first glance they suggest that early-type
galaxies are far more closely related to the group properties than the
late types.  However most of these systems are dominated by
early-types, and only 3 systems have more than half their total light
originating in late-type galaxies.  Given this lower number of spirals
in these groups it is unsurprising that the relations involving them
show more scatter. It should also be noted that the correlations for
$L_B : T$ and $L_X : L_B$ are stronger when all morphological types
are combined. The lack of any correlation in the
$\beta_{\textrm{spec}}$ versus late-type light could also be due to
poorer statistics for the late-type data. The correlations seen
between $\beta_{\textrm{spec}}$ and total or early-type light most
likely represent a trend of lower $\beta_{\textrm{spec}}$ in smaller
systems (smaller systems also tend to contain a lower fraction of
early-types) as discussed in
\S\ref{sec:opt_light_results}.

\subsection{The relationship between the central galaxy and the group}

Another interesting relationship is that between the X-ray properties
of a group central galaxy and the X-ray properties of the group (see
Figure~\ref{fig:cen}). While the relation between central galaxy $L_X$
and $L_B$ appears to be consistent with previous relations derived
over a much larger range of $L_B$, there is clearly a substantial
amount of scatter. This scatter is unsurprising as previous work has
noted that the scatter in $L_X$ can be as much as one or two orders of
magnitude (e.g \citealt{canizares87,eskridge95}). However there does
appear to be a strong correlation between the X-ray luminosities of
the galaxy and group. Given this, and the fact that the optical
luminosity of the central galaxy does not correlate with the group
X-ray luminosity, it appears that the X-ray properties of the central
galaxies are more closely related to the group than to the galaxy
itself, in the sense that the most X-ray overluminous (higher values
of $L_X/L_B$) galaxies are found in the brightest groups. This is
consistent with our previous work \citep{helsdon01} which has shown
that the X-ray properties of the central galaxies are different to
those of other non-central galaxies in groups.

The correlation of central galaxy X-ray luminosity with group X-ray
luminosity appears to be a problem for models in which the central
component is due to a central mass excess associated with a central
galaxy (e.g. see \citealt{makishima01} and references therein). In
this model it is argued that the central X-ray component is made up of
a combination of the central galaxy ISM and an an excess of
intracluster gas due to an additional potential drop caused by the
central galaxy. Thus, in this model, the central emission excess is
entirely due to components associated with the central galaxy. This
suggests that there should be stronger correlations between the
central component properties and the galaxy properties, than between
the central component and the group. This is not the case for our
data.

\begin{figure*}
\hspace{0cm}
\psfig{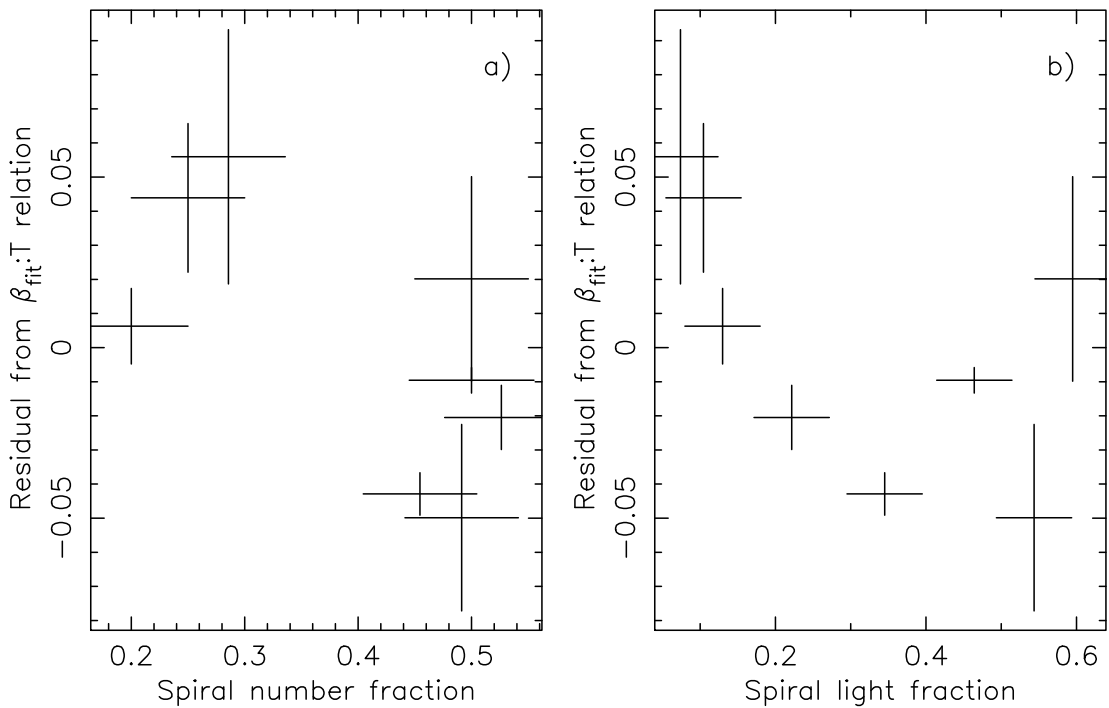}
\caption{\label{fig:btres_spiral}Residual from $\beta_{\mathrm{fit}} : T$ 
  relation (data-model) as a function of a) spiral number fraction and
  b) spiral light fraction, for the eight groups with the most reliable
  values of $\beta_{\mathrm{fit}}$ identified in
  \protect\citet{helsdon00}.}
\end{figure*}

Gas loss from the central galaxies themselves is unable to explain the
high X-ray luminosities or the extent of the emission observed in
these bright central galaxies, and simulations have shown that
additional infalling material is required to adequately reproduce
their observed properties (\citealt{brighenti98},
1999\nocite{brighenti99}). Given the strong correlation between galaxy
and group X-ray luminosities and the fact that almost all of these
central galaxies are in groups which show evidence of a central
temperature drop \citep{helsdon00}, it is likely that this additional
infalling material is in the form of a group cooling flow. If indeed
the X-ray properties of this central component are a property of the
group rather than the galaxy itself, this would account for some of
the observed scatter in the early-type galaxy $L_X/L_B$ relation. A
group cooling flow could also explain the apparent lack of
rotationally enhanced X-ray ellipticity in the cooling flows of
elliptical galaxies \citep{hanlan00}, and the correlation of X-ray
luminosity with the relative sizes of the X-ray and optical emission
\citep{mathews98}, in that the most X-ray overluminous galaxies should
be found in the centre of bigger groups.

We have also seen that the degree of dominance of the central
early-type galaxy does not appear to correlate with any other group
property, including residual from the group $L:T$ relation
(Figure~\ref{fig:ltres_d12}). Previous work on fossil groups
\citep{jones00} has suggested that these groups (which generally have
$\Delta m_{12} \geq 2.5$) may lie above the mean $L:T$ relation.
Fossil groups are thought to be the result of a number of galaxy
mergers within a compact group, thus leaving a single bright
early-type in the centre of the group potential, with no other bright
galaxies nearby.  They would be expected to have high X-ray
luminosities if they formed at an early epoch when the density of the
universe was higher, leading to a higher gas density and thus higher
X-ray luminosity. The number of fossil groups with reliable
temperatures is still small, and the present results do not fit in
with the tentative pattern proposed by \citep{jones00}, despite the
fact that two groups have values of $\Delta m_{12}$ close to that of
fossils (both have $\Delta m_{12} > 2.2$), and may be examples of
local fossils.

\subsection{Galaxy morphology and preheating in groups}
\label{sec:dis:pre}

\begin{figure*}
\hspace{0cm}
\psfig{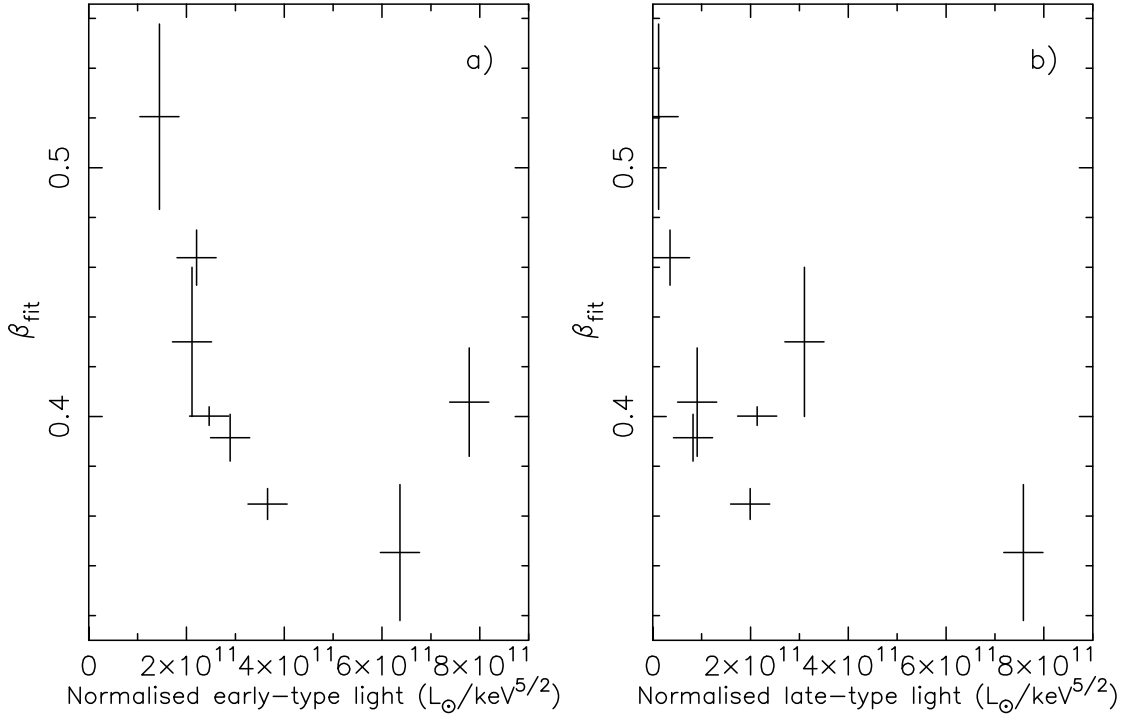}
\caption{\label{fig:scaled_light} a) Early-type and b) late-type light,
 normalised by the total thermal energy of the group as a function of
 $\beta_{\mathrm{fit}}$, for the groups with reliable values of
 $\beta_{\mathrm{fit}}$. Errors in the $x$ direction are arbitrary.}
\end{figure*}

Finally we come to the relation between $\beta_{\mathrm{fit}}$ and
spiral fraction.  Although the sample size is small and the
correlation is weak, Figure~\ref{fig:beta_spiral} suggests that the
systems with higher spiral fractions have flatter profiles. Systems
which have experienced more preheating should show flatter profiles
(e.g. \citealt{voit02}) so this result suggests that spiral galaxies
may play a significant role in any preheating in these systems.  This
is not what would be expected given the results of \citet{arnaud92}
who looked at the correlation between iron mass and optical luminosity
in clusters of galaxies. They found little correlation between iron
mass and spiral luminosity and thus concluded that only early-type
galaxies have contributed to the enrichment of the intracluster
medium.  If early-types are primarily responsible for the enrichment
of the ICM then it would also be expected that they should be the
primary origin of any preheating in these systems, which would produce
trends in the opposite direction to those seen here. However some care
is needed here as groups also show trends in $\beta_{\mathrm{fit}}$
with temperature (and thus mass) in that smaller systems have flatter
profiles. In order to examine the effects of preheating as a function
of morphological type, any trends with system mass must also be taken
into account.

We use two different approaches to attempt to remove the effects of
system mass on $\beta_{\mathrm{fit}}$ -- residuals from the
$\beta_{\mathrm{fit}} : T$ relation, and trends in
$\beta_{\mathrm{fit}}$ and optical light as a function of
morphological type after an appropriate scaling. In addition we can
investigate the morphological origin of preheating by examining the
effect of spiral fraction on the slope of the $L_X : T$ relation. All
three methods are described in more detail below.

\begin{figure*}
\hspace{0cm}
\psfig{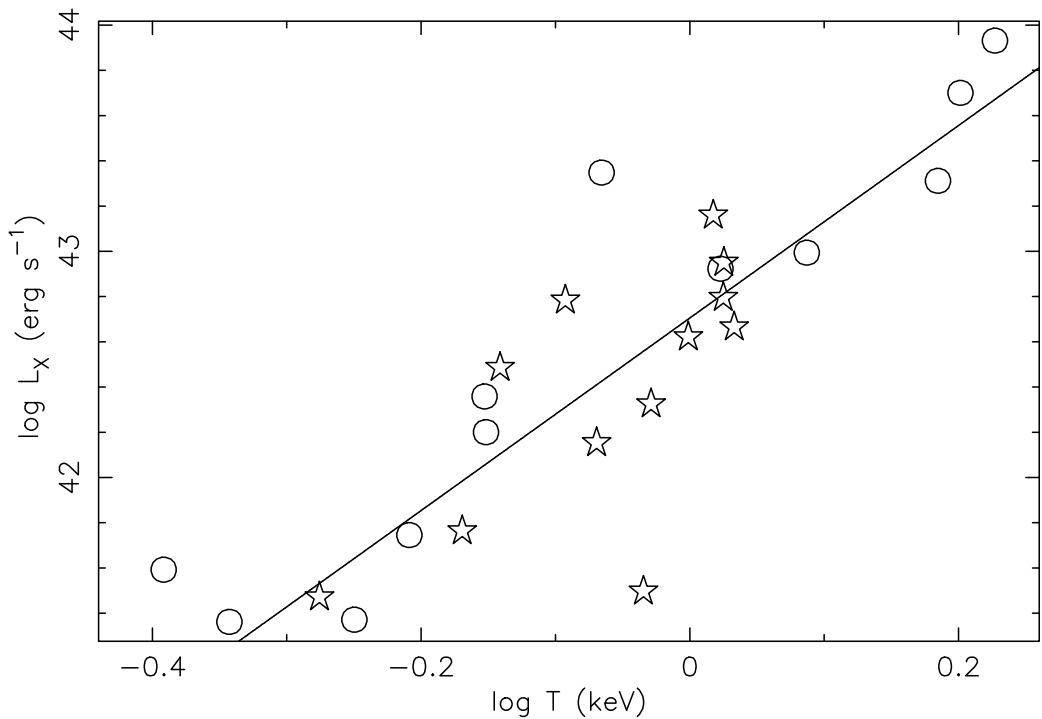}
\caption{\label{fig:ltfr} The $L_X : T$ relation for spiral rich systems
  (stars) and spiral poor systems (circles). The solid line is the best fit 
  $L_X : T$ relation from \protect\citet{helsdon00b}.}
\end{figure*}

The first method examines the residual from the
$\beta_{\mathrm{fit}} : T$ relation as a function of morphological
type. We first fit a mean trend to the cluster and group
$\beta_{\mathrm{fit}}$ and log temperature data given in
\citet{helsdon00}. We then calculate the residual (data - fit) for each of
the eight good quality group points from this trend, and plot it as a
function of spiral fraction.  Do systems which lie above the mean
trend (i.e. have steeper profiles, which would suggest less
preheating) tend to be preferentially elliptical or spiral rich?  The
residuals are plotted against both spiral number fraction and spiral
light fraction in Figure~\ref{fig:btres_spiral}. As can be seen there
are weak trends present ($K=-0.88$ for spiral number fraction, and
$K=-1.73$ for spiral light fraction). Although the sample size is small
and the correlations are not particularly strong, the trends in the
data are in the sense that that systems with higher fractions of
spirals tend to have flatter profiles.

The second method we use to examine the origin of any preheating in
the systems is to look at relations involving optical light for each
morphological type. If early-type galaxies are primarily responsible
for the preheating then the effects of the preheating should scale
with the amount of light in early types. Note that this will also be
true if AGN are primarily responsible for the preheating, as black
hole mass is correlated with the total light of the spheroidal
components of galaxies \citep{magorrian98}. However the effects of any
energy injection will also vary with system mass -- the impact of the
energy injection should scale as the ratio of the injected energy
($\propto L_B$) to the total thermal energy ($M_{\mathrm{gas}}
T$). Given also that $M_{\mathrm{total}} \propto T^{3/2}$ and
$M_{\mathrm{total}}/M_{\mathrm{gas}}$ = constant, we obtain a
scaling factor of $T^{5/2}$ for the total thermal energy of the gas in
self similar systems. Thus if early-type galaxies are the origin of
the preheating in these systems then there should be a correlation
between early-type light divided by $T^{5/2}$ and
$\beta_{\mathrm{fit}}$.

In Figure~\ref{fig:scaled_light} we plot $\beta_{\mathrm{fit}}$ as a
function of both scaled (divided by $T^{5/2}$) early and late-type
light, for the same systems as plotted in
Figures~\ref{fig:beta_spiral} and~\ref{fig:btres_spiral} earlier. As
can be seen there is indeed a correlation of $\beta_{\mathrm{fit}}$
with normalised early-type light ($K=-2.22$, $P=0.0264$).  However there
is also marginal evidence for a correlation with late-type light
($K=-1.73$, $P=0.0836$). This suggests that both early and late-type
galaxies play a role in the preheating of the intragroup medium.

The final method we use to examine the origin of preheating in these
systems is to compare the slope of the $L_X : T$ relation for spiral and
elliptical rich systems. We split the full sample of 24 groups into
two equal sized subsamples based on their spiral fractions. The groups
in the early-type rich subsample all have spiral fractions of
$\leq$0.4. If early-type galaxies are primarily responsible for the
preheating of the IGM then one would expect the spiral rich sample to
have a shallower $L:T$ slope than the spiral poor sample. As can be
seen in Figure~\ref{fig:ltfr}, the two subsamples follow a similar
relation -- the spiral rich subsample has a slope of 5.5 $\pm$ 0.9
while the spiral poor subsample follows a relation of slope 4.3 $\pm$
0.5. Again, this $L_X : T$ data does not support a scenario in which
early-types are the dominant source of any preheating, but suggests
that late-type galaxies play at least a comparable role to
early-types.

The above results favour a situation where spirals have contributed
significantly to the heating, and hence also presumably to the
enrichment, of the intragroup medium, probably at a level comparable
to the contribution from early-types. We can use this in an attempt to
discriminate between AGN heating and supernova-driven wind heating, as
the AGN heating should be proportional to the luminosity of the
spheroidal component whereas the rate of supernovae is roughly
proportional to the total luminosity. Note that, in both cases these
heating estimates are for the systems at the epoch at which they are
observed, and that ideally a hierarchical model, integrated over the
age of the Universe, would be needed to estimate the effects of the
different heating mechanisms. 

If indeed spiral galaxies do play a significant role in the preheating
of the intragroup medium then it is unlikely that AGN heating is the
dominant preheating mechanism, as the most massive AGN are found in
early-type galaxies. However some caution is needed with these results
as the actual stellar mass contribution from disks (as compared to
spheroids) in these groups is fairly small ($\sim$ 10\%-15\% ---
spirals have spheroidal components as well), so only modest trends
with spiral fraction would be expected. However, this result does
suggest that galaxy winds may play the dominant role in the preheating
of the intragroup medium and it also suggests that spiral galaxies may
have played a significant role in the metal enrichment of
intergalactic space.


\section{Conclusions}
\label{sec:conc}

Using optical data drawn from the literature we have examined the
relationship between the X-ray properties of a galaxy group and its member
galaxies. Our main results may be summarised as follows:

\begin{enumerate}
  
\item{The total optical light in an X-ray bright galaxy group appears to correlate well
    with the X-ray temperature. In addition the $L_B : T$ relation
    ($L_B\propto T^{1.6\pm0.2}$) appears to be consistent with a
    straight extrapolation of the cluster relation and is also
    consistent with self similar scaling of galaxy systems. This
    continuation of the cluster trend and the small amount of scatter
    observed about this line suggest that the star formation
    efficiency is fairly constant across all these systems.}
  
\item{A constant star formation efficiency does not support a scenario in
    which cooling of material is responsible for the observed steepening of
    the $L_X : T$ relation. The substantial amount of cooled material
    required to reproduce the $L_X : T$ relation would require the 
    star formation efficiency in groups to be substantially higher than in
    clusters, unless the cooled gas fails to form stars, and is sequestered
    as baryonic dark matter.}
  
\item{The $L_X : L_B$ relation ($L_X\propto L_B^{2.6\pm0.4}$) is
    significantly steeper than has been seen in some previous studies,
    although this steeper slope is consistent with preheating models which
    also steepen the $L_X : T$ relation. In addition, the scatter in $L_X$ in
    both the $L_X : L_B$ and $L_X : T$ relations is much greater than in $L_B$
    or $T$, which suggests that the result of any physical processes in
    these systems (e.g. preheating) primarily affects the distribution of
    the gas in these systems.}
  
\item{The optical light in these groups appears to be more centrally 
    concentrated than the light in clusters. The projected cumulative
    light profile of these groups is given by $L_B(<r) \propto
    r^{0.34\pm0.06}$, which in turn implies a 3D galaxy density
    profile with an index of 2.66. In contrast, with clusters one
    would expect a relation nearer $L_B(<r) \propto r^{1.0}$.}

\item{For the groups with two spatial components to their X-ray emission, 
    the most X-ray overluminous galaxies (high $L_X/L_B$) appear to
    reside in the brightest groups. In addition the X-ray luminosity
    of these central galaxies appears to make up about 25\% that of
    the group itself.  Given that these galaxies are located in the
    centres of groups which show evidence of a central temperature
    drop, it is likely that the X-ray luminosity in these systems is
    primarily a product of the group, such as a group cooling flow. If
    so, then this could explain some of the large scatter seen in the
    $L_X : L_B$ relation for early-type galaxies.}
  
\item{The dominance of the central galaxy as measured by the difference in
    magnitude of the two brightest galaxies does not appear to correlate
    with any of the X-ray properties of the group. Studies of fossil groups
    have suggested that systems with very dominant galaxies may lie above
    the $L_X : T$ relation, however we see no evidence for this. In
    particular, the two groups with the most dominant galaxies lie below
    the $L_X : T$ relation.}
    
\item{We have examined the relationship between galaxy morphology and group 
    properties and see evidence of a weak anti-correlation between
    spiral fraction and group temperature. We have also looked at the
    relation between galaxy morphology and
    $\beta_{\mathrm{fit}}$. $\beta_{\mathrm{fit}}$ should be related
    to the amount of preheating in a system. Although our sample of
    good values of $\beta_{\mathrm{fit}}$ is small we do see a weak
    correlation between spiral fraction and $\beta_{\mathrm{fit}}$ in
    the sense of the profiles being flatter as spiral fraction
    increases. However as $\beta_{\mathrm{fit}}$ also scales with
    system mass we need to also take this into account. We do this in
    two ways. Firstly we examine the residual from the mean
    $\beta_{\mathrm{fit}}:T$ relation as a function of spiral
    fraction. Secondly we scale the optical light (which we assume to
    be proportional to the amount of preheating) as a function of
    galaxy morphology by a factor proportional to the total thermal
    energy of the X-ray gas and look for correlations between this and
    $\beta_{\mathrm{fit}}$. Although the statistics are poor and the
    expected contribution from disks is modest, the results from both
    these studies conflict with a scenario in which early-type
    galaxies alone are responsible for the preheating of the
    intragroup medium.  Instead it appears that spiral galaxies may
    play a comparable role.  This conclusion is also supported by
    examining the slope of the $L_X : T$ relation for spiral rich and
    spiral poor groups.}
  
\item{If indeed spiral galaxies do play a significant role in the
    preheating of galaxy systems then it is unlikely that AGN heating
    is the dominant preheating mechanism, as the most massive AGN are
    found in early-type galaxies (although some caution is needed as
    the stellar mess contribution from disks in these systems is
    small) . This suggests that galaxy winds are likely to play the
    dominant role in any preheating. In addition if spiral galaxies
    are significantly involved in any preheating then they are also
    likely to have contributed to the enrichment of the intra-group
    medium.}

\end{enumerate} 

This study demonstrates the potential of looking at trends, and the scatter
about those trends, in the X-ray and optical properties of galaxy groups.
As more information, such as the metallicity and temperature structure of
these groups, becomes available from the new generation of X-ray
telescopes, it will be possible to look in far greater detail at the
complex relationship between groups and their member galaxies.


\section{Acknowledgements}

The authors would like to thank the referee, Gary Mamon, for his many
useful comments and suggestions, and also Alexis Finoguenov for
further useful comments and discussions.  The data used in this work
have been obtained from the Leicester database and archive service
(LEDAS). This work made use of the Starlink facilities at Birmingham,
the NASA/IPAC Extragalactic Database (NED ---
http://nedwww.ipac.caltech.edu) and the LEDA database
(http://leda.univ-lyon1.fr). SFH acknowledges financial support from
the University of Birmingham.


\bibliography{../bibtex/reffile}
\mbox{}\label{lastpage}

\end{document}